\documentclass[amsmath,amssymb,prc,final,showpacs,twocolumn]{revtex4-1} 

\usepackage{bm,hyphenat,xspace} 
\usepackage{graphicx,epsfig}

\newcommand{\scl}{0.62}

\newcommand{\A}[2]{{}^{#1}\mathrm{#2}}

\newcommand{\clh}{\mathcal{H}}

\newcommand {\mbf}[1]{{\mathbf{#1}}}

\newcommand{\cm}{\mathrm{c\!\:\!.m\!\:\!.}}

\begin{document}

\title {Faddeev-type calculation of three-body nuclear reactions
including core excitation}

\author{A.~Deltuva} 
\affiliation{Centro de F\'{\i}sica Nuclear da Universidade de Lisboa, 
P-1649-003 Lisboa, Portugal }

\received{June 27, 2013}


\begin{abstract}
\begin{description}
\item[Background]
The core excitation, being an  important reaction mechanism, so far
is not properly included in most calculations
of three-body nuclear reactions.
\item[Purpose]
We aim to include the excitation of the core nucleus using an
exact Faddeev-type framework for nuclear reactions in the  three-body 
(core + neutron + proton) system.
\item[Methods]
We employ Alt, Grassberger, and Sandhas (AGS) integral equations for the 
three-particle transition operators and solve them in the momentum-space
framework. The Coulomb interaction is included via the method of screening
and renormalization.
\item[Results]
We calculate elastic, inelastic, and transfer reactions involving
$\A{10}{Be}$ and  $\A{24}{Mg}$ nuclear cores.
\item[Conclusions]
Important effects of the core excitation are found, often
improving the description of the experimental data. 
In the neutron transfer reactions the core excitation effect is by far
not just a simple reduction of the cross section 
by the respective spectroscopic factor.
This indicates that widely used extraction of the spectroscopic factors 
from the ratio of the experimental and theoretical  transfer cross sections
is unreliable approach. 
\end{description}
\end{abstract}

 \maketitle


Direct nuclear reactions of three-body nature 
provide an important test for models of nuclear dynamics.
Widely studied examples are deuteron $(d)$ scattering from
a stable nucleus $(A)$ and proton $(p)$ scattering from a weakly
bound nucleus $(An)$ consisting of a core $A$ and a neutron $(n)$.
Elastic, inelastic, breakup, transfer, and charge-exchange reactions 
can be realized in such nuclear systems.
A number of theoretical approaches has been developed and used for the
analysis of these reactions, ranging from relatively simple
distorted-wave Born approximation (DWBA),
 adiabatic wave approximation (ADWA) \cite{johnson:70a},
 and semiclassical eikonal approximation  \cite{baye:09a} to more sophisticated
continuum discretized coupled channels (CDCC) method \cite{austern:87}
and exact Faddeev-type scattering theory  \cite{faddeev:60a,alt:67a}.
The technical complexity of the latter precluded it from being used widely
for the analysis of nuclear reaction data, but it was namely the 
Faddeev-type method that allowed to test the accuracy of the traditional 
nuclear reaction approaches and revealed their limitations in particular
cases \cite{deltuva:07d,nunes:11b,upadhyay:12a}.

However, the standard technical implementations of the above-mentioned methods 
do not take explicitly into account the excitations of the core nucleus $A$
that may be an important reaction mechanism.
Some of them, like DWBA and ADWA,
assume that the core excitation effects simply lead to the reduction
of transfer cross sections by a spectroscopic factor (SF)
but the reliability of this approximation is unknown.
Nevertheless, in last years several attempts have been made to include
the core excitation into the three-body reaction dynamics.
The extensions of the DWBA \cite{crespo:11a,moro:12a}
and CDCC methods \cite{summers:06a,summers:07a}, the latter called XCDCC, 
were developed to calculate the breakup of one-neutron halo nucleus,
however, with contradicting results.  
Another extension of CDCC with additional approximations
was attempted in elastic and inelastic deuteron scattering \cite{chau:inpc07}.
Faddeev-type theory with separable potentials was proposed in 
Ref.~\cite{mukhamedzhanov:12a},
however, without any numerical calculations.

On the other hand, a similar kind of scattering calculations with dynamic
excitation of involved particles was developed in 90's for the three-nucleon 
($3N$) system \cite{nemoto:98a,nemoto:98b,chmielewski:03a}. 
It includes the excitation of a nucleon into a $\Delta$ isobar via 
coupled-channel potential \cite{hajduk:83a}
in the framework of momentum-space symmetrized
Alt, Grassberger, and Sandhas (AGS) equations  \cite{alt:67a}
that are equivalent to Faddeev equations \cite{faddeev:60a} but are
formulated for the transition operators.
While early calculations \cite{nemoto:98a,nemoto:98b,chmielewski:03a} relied
on a separable representation of the interaction and were restricted to
neutron-deuteron scattering, later developments include fully realistic
non-separable coupled-channel potential with $\Delta$ isobar excitation
\cite{deltuva:03a,deltuva:03c} and Coulomb interaction 
\cite{deltuva:05a} 
using the method of screening and renormalization \cite{taylor:74a,alt:80a}.
The latter achievement allowed the application of the AGS equations also
to three-body nuclear reactions in the $A+n+p$ system 
\cite{deltuva:07d,deltuva:09a,deltuva:09b},
albeit considering $A$ as an inert particle so far.
It is the aim of the present work to overcome this limitation by developing
exact Faddeev-type calculations for reactions in the  $A+n+p$ system 
where the core nucleus $A$ with mass $m_A$ can be excited into $A^*$
 with mass $m_{A^*}$ when interacting with nucleons; 
for simplicity we consider here only one excited state $A^*$ but the
generalization to several excitations is straightforward.
For this purpose we combine the strategy of Refs.
\cite{deltuva:03a,deltuva:03c} and \cite{deltuva:07d,deltuva:09a,deltuva:09b}.
The developed method will be illustrated by the results for elastic, inelastic, 
and transfer reactions involving $\A{10}{Be}$ and  $\A{24}{Mg}$ nuclear cores.

The three-particle scattering problem is formulated in the Hilbert space 
 $\clh_g \oplus \clh_x$ with two sectors, where $\clh_g$ 
contains states for $A+n+p$ free relative motion 
$| \mbf{p}_\alpha  \mbf{q}_\alpha \rangle_g $
and $\clh_x$ contains free  $A^*+n+p$ states 
$| \mbf{p}_\alpha  \mbf{q}_\alpha \rangle_x $.
Here $\mbf{p}_\alpha$ and  $\mbf{q}_\alpha$ are the three-particle
Jacobi momenta in any of three possible configurations with
spectator particle $\alpha$; the respective reduced masses will be denoted
by $\mu_\alpha$ and  $M_\alpha$. For brevity we suppress
the dependence on spin quantum numbers.
The extended free Hamiltonian $H_0$ in addition to kinetic energy
operator includes also the intrinsic Hamiltonian of the core
\cite{moro:12a,mukhamedzhanov:12a} whose contribution
is the mass difference relative to the $A+n+p$ threshold, i.e.,
$ H_0| \mbf{p}_\alpha  \mbf{q}_\alpha \rangle_g =
(p_\alpha^2/2\mu_\alpha + q_\alpha^2/2M_\alpha)
| \mbf{p}_\alpha  \mbf{q}_\alpha \rangle_g  $ and 
$ H_0| \mbf{p}_\alpha  \mbf{q}_\alpha \rangle_x =
(p_\alpha^2/2\mu_\alpha + q_\alpha^2/2M_\alpha + \Delta m)
| \mbf{p}_\alpha  \mbf{q}_\alpha \rangle_x  $ 
with $\Delta m = m_{A^*} - m_{A}$.
The two sectors  $\clh_g$ and  $\clh_x$ are coupled by the core-nucleon
interaction. 

In such an extended Hilbert space the AGS scattering equations
for three-particle transition operators $U_{\beta\alpha}$
have the standard form
\begin{equation}  \label{eq:Uba}
U_{\beta \alpha}  = \bar{\delta}_{\beta\alpha} \, G^{-1}_{0}  +
\sum_{\gamma=1}^3   \bar{\delta}_{\beta \gamma} \, T_{\gamma} 
\, G_{0} U_{\gamma \alpha}.
\end{equation}
Here  $ \bar{\delta}_{\beta\alpha} = 1 - \delta_{\beta\alpha}$ and
$G_0 = (E+i0-H_0)^{-1}$ is the free resolvent at the 
available three-particle energy $E$ in the center of mass (c.m.) system;
obviously, $G_0$ is diagonal in the two Hilbert sectors.
The  two-particle transition operator 
\begin{equation}  \label{eq:T}
T_{\gamma} = v_{\gamma} + v_{\gamma} G_0 T_{\gamma}
\end{equation}
for the pair $\gamma$ in the odd-man-out notation is derived from
the respective two-particle potential $v_{\gamma}$;
unless  $\gamma$ corresponds to $np$ pair,  $v_{\gamma}$ and $T_{\gamma}$
couple the sectors  $\clh_g$ and  $\clh_x$.
Obviously, the three-particle transition operators $U_{\beta\alpha}$
couple the two sectors too; their on-shell matrix elements 
$\langle \phi_\beta |U_{\beta\alpha} | \phi_\alpha \rangle$
between the respective asymptotic channel states 
are scattering amplitudes from which observables for all reactions
allowed by the chosen Hamiltonian can be calculated.
The two-cluster channel states $| \phi_\alpha \rangle$
with $\alpha=1,2,3$ are given by the respective two-particle
bound state wave function times the plane wave for the
relative spectator-pair motion; the former has components
in both sectors $\clh_g$ and  $\clh_x$ if $\alpha$ corresponds to
the core-nucleon pair.

The original AGS equations are formulated for short-range 
 potentials $v_{\gamma}$. However, as already mentioned, the long-range
repulsive Coulomb force acting within one pair of particles
can be included using  the method of screening and renormalization 
\cite{taylor:74a,alt:80a} in the AGS framework.
The long-range part of the scattering amplitude is known analytically,
whereas the Coulomb-distorted short-range part 
 is obtained by solving  the AGS equations 
with nuclear plus screened Coulomb potentials numerically
and ensuring the convergence of the results with the screening radius.
An efficient way to achieve this goal is proposed
 in Ref.~\cite{deltuva:05a}.
The solution technique relies on a partial-wave decomposition 
and discretization of integrals using Gaussian quadratures
with special (standard) weights for singular (non-singular) integrands.  
Thus, while the numerical techniques for solving the
AGS equations with core excitation are taken over from
Refs.~\cite{deltuva:03a,deltuva:03c,deltuva:05a,deltuva:07d,deltuva:09a},
there are two novel aspects vis-a-vis those previous calculations
relevant for practical implementation:
i) The core excitation energy  $\Delta m$, being 3.368 MeV for
$\A{10}{Be}$ and 1.369 MeV for $\A{24}{Mg}$,
is much smaller as compared to the $\Delta $-nucleon mass difference 
of nearly 300 MeV.
Thus, while in Refs.~\cite{deltuva:03a,deltuva:03c}
the  $\Delta$-isobar excitations are virtual with
no $\Delta$ components in asymptotic channel states, the channels
with the core in both ground $A$ and excited  $A^*$ states are
open in the considered energy regime.
This implies also the presence of singularities in the momentum-space
integral equation kernels in both sectors $\clh_g$ and  $\clh_x$;
they are treated by the real axis integration as described in detail
in Refs.~\cite{chmielewski:03a,deltuva:phd}.
ii) Typically, the core in the excited state has higher spin.
In the examples of the present work, $\A{10}{Be}$ and  $\A{24}{Mg}$, 
 the spin/parity is
$0^+$ and $2^+$ for $A$ and $A^*$, respectively. In this case the number
of partial waves to be included in the sector  $\clh_x$ is larger by a factor
of 5 as compared to  $\clh_g$, leading to increase of computer time
by a factor of 25. Thus, when including the core excitation 
the parallelization of calculations becomes mandatory. 

Finally, we note that the present formulation of three-body reactions
with core excitation is equivalent to the one of Ref.~\cite{mukhamedzhanov:12a}
except for a different way to include Coulomb force and no need
for separable potentials.

The dynamic input to AGS equations are the potentials $v_{\gamma}$ for all 
three pairs. As the $np$ interaction we take the realistic
CD Bonn potential \cite{machleidt:01a}, in contrast to  a simple Gaussian  
$np$ potential used in  DWBA and XCDCC calculations \cite{moro:12a,summers:07a}.
The nucleon-core potentials are less constrained having many possible choices.
In this work we do not aim to use the most sophisticated model but rather 
to isolate the effects of the core excitation. We therefore  perform
calculations with several models of $nA$ and $pA$ interactions:

a) The single particle (SP) model neglects the core excitation 
and uses the original 
Chapel Hill 89 (CH89) \cite{CH89} optical potential (OP) for the
$pA$ pair and for the $nA$ pair in the partial waves without bound states.
The  $nA$ potential in the partial waves with bound states is 
real with the strength adjusted to binding energy.
As the OP parameters are energy dependent, various choices are possible
 for the energies at which $v_{NA}$ are to be calculated. 
For the $\A{10}{Be}+n+p$ system where we are mostly interested in 
neutron transfer reactions
we choose to fix  $v_{pA}$  at the proton energy in the $p+\A{11}{Be}$
channel and $v_{nA}$  at half the deuteron energy in the $d+\A{10}{Be}$
channel. This is exactly the choice labeled FADD in Ref.~\cite{nunes:11b};
the $n$-$\A{10}{Be}$ binding potential is taken over from Ref.~\cite{nunes:11b}
as well.
For $d+\A{24}{Mg}$ elastic and inelastic scattering the CH89 OP
taken at half the initial deuteron energy  is
used for both $pA$ and  $nA$  pairs in all partial waves, i.e.,
the $\A{25}{Mg}$ bound states are not supported.
This choice corresponds to the one labeled FAGS in  Ref.~\cite{upadhyay:12a}.
The respective strategies in fixing the OP parameters 
will be used also with other nucleon-core interaction models
described below.

b) As in Refs.~\cite{crespo:11a,moro:12a,summers:07a},
the $nA$ and $pA$ interactions with the 
core excitation (CX) are obtained from the rotational model
for the core \cite{nunes:96a}.  It assumes  that the
core has a permanent quadrupole deformation characterized
by the deformation parameter $\beta_2$.
The radial  dependence of the nucleon-core potential is given by the
Woods-Saxon function $f(r,R,a) = \{1+\exp[(r-R)/a]\}^{-1}$ and,
eventually, by its derivative,
with the radius $R = R_0(1+\beta_2 Y_{20}(\hat{\xi}))$ 
depending on the quadrupole deformation and the internal core degrees of
freedom $\hat{\xi}$ in the body-fixed frame. 
Obviously, this approach is equivalent to the replacement
of the undeformed central potential $v_{NA}(r)$ by 
$v_{NA}(r-\delta_2Y_{20}(\hat{\xi}))$ where $\delta_2 = \beta_2 R_0$ is the
deformation length. The resulting noncentral 
 potential is expanded into multipoles
as in Refs.~\cite{crespo:11a,moro:12a} with the $\lambda=2$ multipole term being
responsible for the core excitation/deexcitation. 
In addition we include also the standard undeformed spin-orbit potential
as provided by the CH89 parametrization \cite{CH89}.
The geometry parameters for the
$n$-$\A{10}{Be}$ binding potential in the $1/2^+$ state
are taken over from Ref.~\cite{nunes:96a},
i.e., $R_0 = 2.483$ fm, $a = 0.65$ fm, and  $\beta_2 = 0.67$,
resulting  $\delta_2 = 1.664$ fm. The strengths of the central 
$V_0 = -54.45$ MeV and spin-orbit $V_{so} = -8.5 \, \mathrm{MeV\,fm^2}$
parts are slightly readjusted to reproduce better
the  $\A{11}{Be}$ ground state  energy of $-0.504$ MeV.
Its wave-function component with $s$-wave neutron coupled to the
ground state $0^+$ of $\A{10}{Be}$ has the weight (SF) of 0.855. 
The OP's for the $nA$ pair in other partial waves and for $pA$ pair
are obtained by deforming the central part
of the CH89 potential with $\delta_2 = 1.664$ fm for  $\A{10}{Be}$.
For  $\A{24}{Mg}$ the values of $\beta_2$ range from 0.35 to 0.6  \cite{mg:d};
we therefore perform  two CX calculations with
$\beta_2 = 0.4$ and 0.6,  corresponding to  $\delta_2 = 1.352$ and 2.028 fm.
In all cases 
the potentials are supplemented by the undeformed spin-orbit part of CH89.

c) To separate the effects of the core excitation due to interaction with
the neutron and proton, for reactions involving  $\A{10}{Be}$
we perform also the calculations with hybrid model CXn, that
uses the $pA$ potential from the SP model but  the
 $nA$ potential from the CX model.

\begin{figure}[!]
\begin{center}
\includegraphics[scale=0.56]{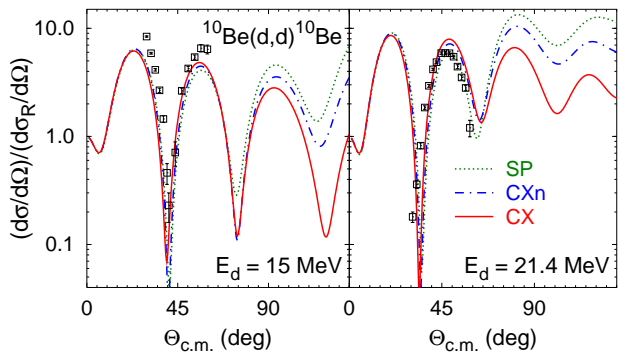}
\end{center}
\caption{\label{fig:bede}  (Color online)
Differential cross section 
for  $d + \A{10}{Be}$ elastic scattering at $E_d = 15$ and 21.4 MeV.
Results including core excitation in both $n$-$\A{10}{Be}$
and $p$-$\A{10}{Be}$ interactions (solid curves), only in 
$n$-$\A{10}{Be}$ interaction (dashed-dotted curves), and neglecting 
the core excitation (dotted curves) are compared with 
experimental data from Ref.~\cite{dBe12-21}.}
\end{figure}

Results obtained with these interaction models
for the $d + \A{10}{Be}$ elastic scattering 
at $E_d = 15$ and 21.4 MeV deuteron lab energy are presented in
Fig.~\ref{fig:bede}. The differential  cross section 
$d\sigma/d\Omega$ divided by the 
Rutherford cross section $d\sigma_R/d\Omega$
is shown as a function of the c.m. scattering angle $\Theta_\cm$. 
The effect of the core excitation is most
pronounced at large scattering angles $\Theta_\cm > 80^\circ$ where
it reduces the  cross section by a factor of 2 to 5.
In contrast, around $\Theta_\cm = 50^\circ$ where the experimental
data from Ref.~\cite{dBe12-21} are available, the core excitation
slightly increases the cross section improving the agreement with data at
 $E_d = 15$ but worsening at 21.4 MeV. The predictions of the CXn
model usually lie between those of SP and CX, indicating that
the core excitation effects due to $n$-$\A{10}{Be}$
and $p$-$\A{10}{Be}$ interactions are quite similar for the
elastic scattering of deuteron.

\begin{figure}[!]
\begin{center}
\includegraphics[scale=\scl]{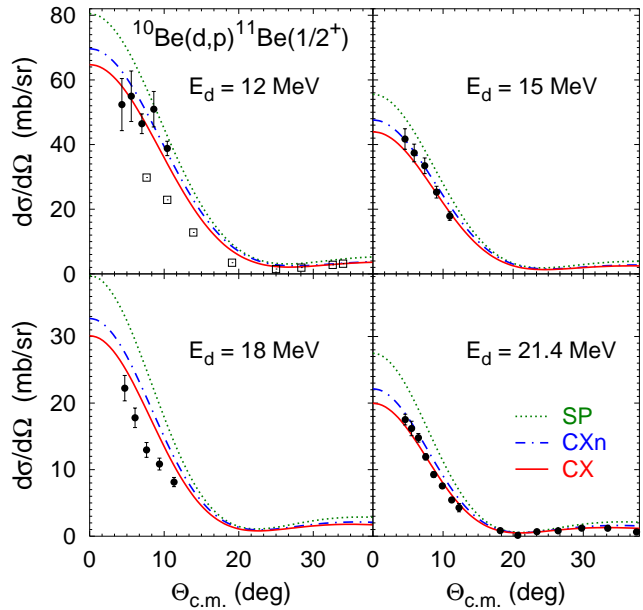}
\end{center}
\caption{\label{fig:bedps}  (Color online)
Differential cross section for the transfer reaction  
$\A{10}{Be}(d,p)\A{11}{Be}$  at $E_d = 12$, 15, 18, and 21.4 MeV
leading to the ground state $1/2^+$ of $\A{11}{Be}$.
Curves as in Fig.~\ref{fig:bede} and 
experimental data from Refs.~\cite{dBe12-21,dBe12p}.}
\end{figure}

The differential  cross section for the deuteron stripping reaction 
$\A{10}{Be}(d,p)\A{11}{Be}$  
leading to the ground state $1/2^+$ of $\A{11}{Be}$ is shown
in  Fig.~\ref{fig:bedps} for $E_d = 12$, 15, 18, and 21.4 MeV.
The most important core excitation effect is observed at small
scattering angles where it decreases the  cross section.
Here the contribution of the core excitation in the 
 $n$-$\A{10}{Be}$ pair is more sizable than  in the $p$-$\A{10}{Be}$ pair.
With increasing energy the  differential  cross section decreases but
the relative effect of the core excitation increases,
leading to quite good agreement between CXn and CX predictions
and experimental data from Ref.~\cite{dBe12-21},
except at $E_d = 18$ MeV. Note, however, that theoretical
results vary smoothly with energy while experimental data exhibit
an abrupt decrease from  $E_d = 15$ to 18 MeV.
The data at  $E_d = 12$ MeV from  Refs.~\cite{dBe12-21}
and \cite{dBe12p} also seem to be inconsistent.

\begin{figure}[!]
\begin{center}
\includegraphics[scale=0.56]{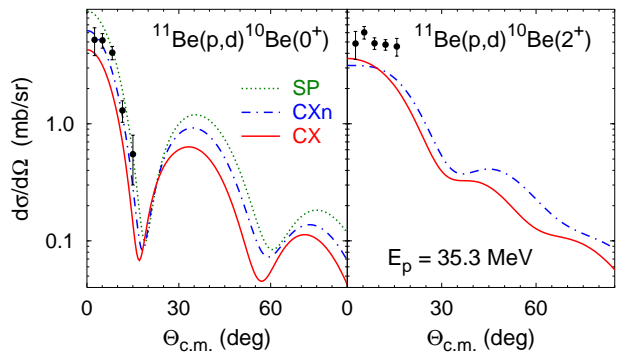}
\end{center}
\caption{\label{fig:bepd}  (Color online)
Differential cross section for the transfer reaction  
$\A{11}{Be}(p,d)\A{10}{Be}$  at $E_p = 35.3$ MeV
leading to the ground ($0^+$) and excited ($2^+$) states of $\A{10}{Be}$.
Curves as in Fig.~\ref{fig:bede} and 
experimental data from Ref.~\cite{winfield:01}.}
\end{figure}

Consistently with the observed energy dependence,
even larger core excitation effect is seen in 
Fig.~\ref{fig:bepd} for the differential  cross section of the
deuteron pickup reaction $\A{11}{Be}(p,d)\A{10}{Be}$  
at proton lab energy $E_p = 35.3$ MeV, corresponding to
$E_d = 40.9$ MeV in the time-reverse reaction $\A{10}{Be}(d,p)\A{11}{Be}$.
Furthermore, in Fig.~\ref{fig:bepd} we present also
the the differential  cross section for the 
 reaction $\A{11}{Be}(p,d)\A{10}{Be}$  leading to the $\A{10}{Be}$
excited state $2^+$ that is inaccessible in the calculations
neglecting the  core excitation. For both reactions
in Fig.~\ref{fig:bepd} the experimental data from Ref.~\cite{winfield:01}
are  underpredicted by CX and CXn  calculations.
Simultaneously we calculated also $p+\A{11}{Be}$ elastic scattering 
(not shown) and found the core excitation effect to be insignificant. 

\begin{figure}[!]
\begin{center}
\includegraphics[scale=0.56]{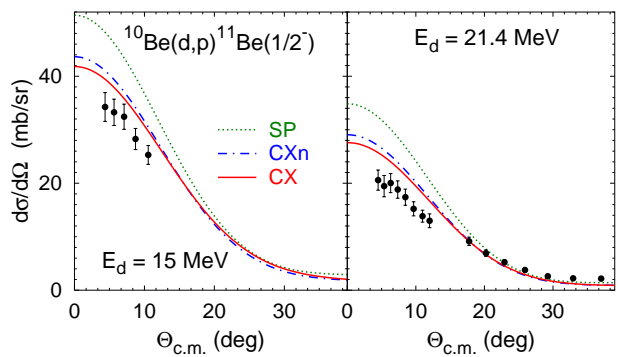}
\end{center}
\caption{\label{fig:bedpp}  (Color online)
Differential cross section for the transfer reaction  
$\A{10}{Be}(d,p)\A{11}{Be}$  at $E_d = 15$ and 21.4 MeV
leading to the excited state ($1/2^-$) of $\A{11}{Be}$.
Curves as in Fig.~\ref{fig:bede} and 
experimental data from Refs.~\cite{dBe12p}.}
\end{figure}

The differential  cross section for the neutron transfer reaction 
$\A{10}{Be}(d,p)\A{11}{Be}$  leading to $\A{11}{Be}$ in its excited state 
$1/2^-$ is shown
in  Fig.~\ref{fig:bedpp} for $E_d = 15$ and 21.4 MeV.
In this case  a real $n$-$\A{10}{Be}$ potential was used also in the 
$1/2^-$ state with the strength $V_0 = -49.62$ MeV resulting an excited
bound state with the energy of $-0.184$ MeV and SF  of 0.786 for
the component with $p$-wave neutron coupled to the
ground state of $\A{10}{Be}$.
Much like in Fig.~\ref{fig:bedps},
the core excitation effect is mostly pronounced
 at small scattering angles and decreases the  cross section. 
It is even more strongly dominated by the $n$-$\A{10}{Be}$ pair.
However, unlike in  Fig.~\ref{fig:bedps},
 the experimental data from Refs.~\cite{dBe12-21,dBe12p} 
are overpredicted by the calculations.

\begin{figure}[!]
\begin{center}
\includegraphics[scale=0.55]{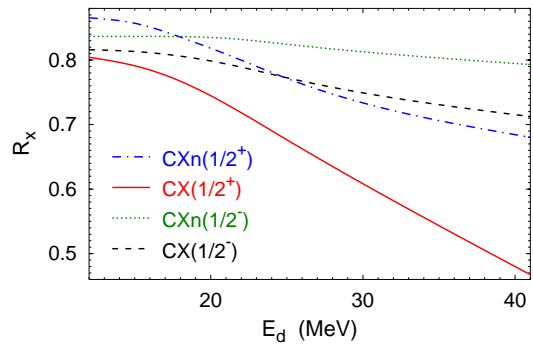}
\end{center}
\caption{\label{fig:r}  (Color online)
Ratios $R_{x}$ of differential cross sections for 
the transfer reactions  $\A{10}{Be}(d,p)\A{11}{Be}$  
calculated with and without core excitation
for  $1/2^+$ and $1/2^-$ states of $\A{11}{Be}$.}
\end{figure}

In Fig.~\ref{fig:r} we  show the core excitation effect
in the neutron transfer reactions on a finer scale.
We observe that the ratios 
$R_{x} = (d\sigma/d\Omega)_{x}/(d\sigma/d\Omega)_{\rm SP}$, $x$ being
either CX or CXn,
depend only weakly on $\Theta_\cm$ in the angular regime below the 
first minimum. We therefore take these ratios at 
$\Theta_\cm = 0^\circ$ and study their energy dependence 
in Fig.~\ref{fig:r}. In the naive DWBA/ADWA picture 
one could expect these ratios to be equal to the respective
SF for the $n$-$\A{10}{Be}$ bound state wave-function
component with the core in its ground state $0^+$, i.e.,
0.855 (0.786) for transfer to $\A{11}{Be}$ ground (excited) state.
Obviously, in a more sophisticated theory there are deviations
from this naive picture as Fig.~\ref{fig:r} demonstrates.
For the $\A{11}{Be}$ excited state $1/2^-$ 
these deviations are moderate, up to 10\%.
However, for the  transfer to the  $1/2^+$ ground state of $\A{11}{Be}$ 
the ratios $R_{\rm CX}$ and $R_{\rm CXn}$ decrease
with increasing energy and, especially $R_{\rm CX}$, significantly
deviate from the SF value of 0.855, reaching 0.47 at $E_d = 41$ MeV.
Thus, these results indicate that extracting the SF from the ratio
of the experimental and theoretical  transfer cross sections
as usually done in DWBA/ADWA is, in general, an unreliable
approach, although in particular cases like
$\A{10}{Be}(d,p)\A{11}{Be}(1/2^-)$ at low energies 
it may provide quite reasonable results.
We emphasize that the present Faddeev-type reaction theory
does not allow for a direct extraction of SF but reveals
to what extent the employed dynamic model with core excitation
predicting particular values for SF and reaction observables
is compatible with given experimental data.

\begin{figure}[!]
\begin{center}
\includegraphics[scale=0.58]{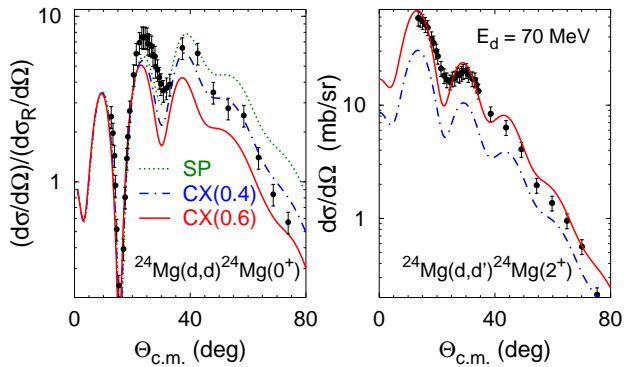}
\end{center}
\caption{\label{fig:mg}  (Color online)
Differential cross section for  $d + \A{24}{Mg}$ elastic (left)
and inelastic (right) scattering at $E_d = 70$ MeV.
Results including core excitation with $\beta_2 = 0.4$
 (dashed-dotted curves) and  $\beta_2 = 0.6$ (solid curves) and neglecting 
the core excitation (dotted curves) are compared with 
experimental data from Ref.~\cite{mg:d}.}
\end{figure}

Finally we present results for elastic
and inelastic scattering of deuterons on  $\A{24}{Mg}$;
the latter reaction is only accessible in the 
dynamic model including core excitation.
In Fig.~\ref{fig:mg} we show the respective 
differential cross sections at  $E_d = 70$ MeV.
In elastic scattering the core excitation effect is most
visible at large angles where it decreases the cross section,
improving the description of the experimental data from  Ref.~\cite{mg:d}
in the case of  $\beta_2 = 0.4$.
However, around the maximum at $\Theta_\cm = 50^\circ$ all 
calculations are below the data.
The inelastic cross section data is underpredicted by theory nearly
by a factor of 2 if one uses $\beta_2 = 0.4$ but is quite well described
using  $\beta_2 = 0.6$.
The shape of the angular dependence with several maxima and minima is 
 reproduced  in both cases.

In summary, we performed calculations of elastic, inelastic, and 
transfer reactions in the nuclear three-body system $A+p+n$
including the excitation of the core nucleus $A$.
Exact scattering equations in
the AGS form were solved with the Coulomb interaction included via
the method of screening and renormalization.
Example results for $\A{10}{Be}$ and  $\A{24}{Mg}$ cores were presented.
Important effects of the core excitation were found, in most cases
improving the description of the experimental data. Furthermore, we
demonstrated that in the neutron transfer reactions the core excitation effect 
cannot be simply simulated by the reduction of the cross section 
according to the respective SF as assumed in DWBA/ADWA calculations.
This deviation from  DWBA/ADWA is not surprising given the fact
that the employed  Faddeev-type reaction theory is an exact one and includes
interactions between the three involved particles up to all orders.
This finding also indicates that extracting the SF from the ratio
of experimental and theoretical  transfer cross sections
is unreliable approach. 
The present Faddeev-type reaction theory
does not allow for a direct extraction of SF but enables a rigorous
test of the employed dynamic model. In this respect further improvements
are possible, e.g., readjusting the parameters of the deformed
optical potentials to get a better description of the two-body data, and
using nonlocal potentials with  the core excitation
since the nonlocality is known to be important in transfer reactions
\cite{deltuva:09b,deltuva:09d}.
We hope that the present work, demonstrating the feasibility of 
exact calculations with core excitation and its importance,
 will stimulate the development of more sophisticated
and precise interaction models.

\vspace{1mm}
The author thanks A.~M.~Moro for discussions and comparison of two-body 
results.



\begin{thebibliography}{10}

\bibitem{johnson:70a}
R.~C. Johnson and P.~J.~R. Soper, Phys. Rev. C {\bf 1},  976  (1970).

\bibitem{baye:09a}
D. Baye {\it et~al.}, 
Phys. Rev. C {\bf 79},  024607  (2009).

\bibitem{austern:87}
N. Austern {\it et~al.}, Phys. Rep. {\bf 154},  125  (1987).

\bibitem{faddeev:60a}
L.~D. Faddeev, Zh.~Eksp.~Teor.~Fiz. {\bf 39},  1459  (1960).

\bibitem{alt:67a}
E.~O. Alt, P. Grassberger, and W. Sandhas, Nucl.~Phys. {\bf B2},  167  (1967).

\bibitem{deltuva:07d}
A. Deltuva {\it et~al.}, Phys.~Rev.~C {\bf 76},  064602  (2007).

\bibitem{nunes:11b}
F.~M. Nunes and A. Deltuva, Phys.~Rev.~C {\bf 84},  034607  (2011).

\bibitem{upadhyay:12a}
N.~J. Upadhyay, A. Deltuva, and F.~M. Nunes, Phys.~Rev.~C {\bf 85},  054621
  (2012).

\bibitem{crespo:11a}
R. Crespo, A. Deltuva, and A.~M. Moro, Phys. Rev. C {\bf 83},  044622  (2011).

\bibitem{moro:12a}
A.~M. Moro and R. Crespo, Phys. Rev. C {\bf 85},  054613  (2012).

\bibitem{summers:06a}
N.~C. Summers, F.~M. Nunes, and I.~J. Thompson, Phys. Rev. C {\bf 74},  014606
  (2006).

\bibitem{summers:07a}
N.~C. Summers and F.~M. Nunes, Phys. Rev. C {\bf 76},  014611  (2007).

\bibitem{chau:inpc07}
P. Chau~Huu-Tai,  in {\em Proceedings of INPC07} (Elsevier, Nucl. Phys. A,
  2008), Vol.~II, p.\ 483.

\bibitem{mukhamedzhanov:12a}
A.~M. Mukhamedzhanov, V. Eremenko, and A.~I. Sattarov, Phys. Rev. C {\bf 86},
  034001  (2012).

\bibitem{nemoto:98a}
S. Nemoto {\it et~al.}, Few-Body Systems {\bf 24},  213  (1998).

\bibitem{nemoto:98b}
S. Nemoto {\it et~al.}, Few-Body Systems {\bf 24},  241  (1998).

\bibitem{chmielewski:03a}
K. Chmielewski {\it et~al.}, Phys.~Rev.~C {\bf 67},  014002  (2003).

\bibitem{hajduk:83a}
C. Hajduk, P.~U. Sauer, and W. Strueve, Nucl.\@ Phys.\@ {\bf A405},  581
  (1983).

\bibitem{deltuva:03a}
A. Deltuva, K. Chmielewski, and P.~U. Sauer, Phys.~Rev.~C {\bf 67},  034001
  (2003).

\bibitem{deltuva:03c}
A. Deltuva, R. Machleidt, and P.~U. Sauer, Phys.~Rev.~C {\bf 68},  024005
  (2003).

\bibitem{deltuva:05a}
A. Deltuva, A.~C. Fonseca, and P.~U. Sauer, Phys.~Rev.~C {\bf 71},  054005
  (2005); {\bf 72},  054004 (2005).

\bibitem{taylor:74a}
J.~R. Taylor, Nuovo Cimento B {\bf 23},  313  (1974); M.~D. Semon and J.~R.
  Taylor, Nuovo Cimento A {\bf 26}, 48 (1975).

\bibitem{alt:80a}
E.~O. Alt and W. Sandhas, Phys.~Rev.~C {\bf 21},  1733  (1980).

\bibitem{deltuva:09a}
A. Deltuva and A.~C. Fonseca, Phys.~Rev.~C {\bf 79},  014606  (2009).

\bibitem{deltuva:09b}
A. Deltuva, Phys.~Rev.~C {\bf 79},  021602(R)  (2009).

\bibitem{deltuva:phd}
A. Deltuva, Ph.D. thesis, University of Hannover, 2003,
  http://edok01.tib.uni-hannover.de/edoks/e01dh03/374454701.pdf.

\bibitem{machleidt:01a}
R. Machleidt, Phys.~Rev.~C {\bf 63},  024001  (2001).

\bibitem{CH89}
R.~L. Varner {\it et~al.}, Phys. Rep. {\bf 201},  57  (1991).

\bibitem{nunes:96a}
F. Nunes {\it et~al.}, Nucl. Phys. A {\bf 609},  43   (1996).

\bibitem{mg:d}
A. Kiss {\it et~al.}, Nucl. Phys. A {\bf 262},  1   (1976).

\bibitem{dBe12-21}
K.~T. Schmitt {\it et~al.}, Phys. Rev. Lett. {\bf 108},  192701  (2012).

\bibitem{dBe12p}
D.~L. Auton, Nucl.~Phys. {\bf A157},  305  (1970).

\bibitem{winfield:01}
J. Winfield~{\it et al.}, Nucl.~Phys. {\bf A683},  48  (2001).

\bibitem{deltuva:09d}
A. Deltuva, Phys.~Rev.~C {\bf 79},  054603  (2009).

\end{thebibliography}

\end{document}